\documentclass{Rinton-P9x6}

\usepackage{amsmath}
\usepackage{amssymb}
\usepackage{latexsym}
\usepackage{revsymb}

\newtheorem{defi}{Definition}
\newtheorem{lemma}[defi]{Lemma}
\newtheorem{thm}[defi]{Theorem}

\newtheorem{rem}[defi]{Remark}
\newtheorem{prop}[defi]{Proposition}

\newcommand{\qed}{\hfill $\Box$}
\newcommand{\tr}{{\operatorname{Tr}}}
\newcommand{\id}{{\operatorname{id}}}
\newcommand{\supp}{{\operatorname{supp}\,}}
\newcommand{\rank}{{\operatorname{rank}\,}}
\newcommand{\conv}{{\operatorname{conv}}}
\newcommand{\bra}[1]{{\langle{#1}|}}
\newcommand{\ket}[1]{{|{#1}\rangle}}
\newcommand{\ketbra}[1]{{\ket{#1}\!\bra{#1}}}
\newcommand{\C}{{\mathbb{C}}}

\newcommand{\1}{{\openone}}

\newlength{\blank}
\newlength{\equalsign}
\newenvironment{beweis}[1][{\hspace{-\blank}}]{{\noindent\emph{Proof~{#1}.\ }}}{\hfill $\Box$\vskip 0.5\baselineskip}

\begin{document}

\title{Identification via Quantum Channels in the\protect\\
       Presence of Prior Correlation and Feedback}

\author{Andreas Winter}
\address{School of Mathematics, University of Bristol,\\
University Walk, Bristol BS8 1TW, U.K.\\
Email: {\tt a.j.winter@bris.ac.uk}}

\maketitle

\abstracts{Continuing our earlier work ({\tt quant-ph/0401060}),
  we give two alternative
  proofs of the result that a noiseless qubit channel has identification
  capacity $2$: the first is direct by a ``maximal code with random
  extension'' argument, the second is by showing that $1$ bit of
  entanglement (which can be generated by transmitting $1$ qubit)
  and negligible (quantum) communication has identification capacity $2$.
  This generalises a random hashing construction of Ahlswede and Dueck: that $1$
  shared random bit together with negligible communication has identification
  capacity $1$.
  \protect\\
  We then apply these results to prove capacity formulas for various quantum
  feedback channels: passive classical feedback for quantum--classical
  channels, a feedback model for classical--quantum channels, and
  ``coherent feedback'' for general channels.
}

\section{Introduction}
\label{sec:intro}
While the theory of identification via noisy channels\cite{AD1,AD2}
has generated significant interest within the information theory
community (the areas of, for instance, common randomness,\cite{AC}
channel resolvability\cite{han:verdu} and watermarking\cite{steinberg:merhav}
were either developed in response or were discovered to have close connections
to identification),
the analogous theory where one uses
a \emph{quantum channel} has received comparably little attention:
the only works extant at the time of writing are
L\"ober's starting of the theory,\cite{Loeber}
a strong converse for discrete memoryless classical-quantum channels
by Ahlswede and Winter,\cite{AW} and a recent paper by the
present author.\cite{winter:ID}
\par
This situation may have arisen from a perception that such a theory
would not be very different from the classical identification theory,
as indeed classical message transmission via quantum channels,
at a fundamental mathematical level, does not deviate much from its classical
counterpart:\cite{Holevo98,SW97,Ogawa:Nagaoka,Winter99}
coding theorem and converses are ``just like'' in Shannon's classical
channel coding theory, with Holevo information playing the role of
Shannon's mutual information.
(Though we have to acknowledge that it took quite a while before
this was understood, and that there are tantalising differences in
detail, e.g. additivity problems.\cite{shor:add})
\par
In our recent work,\cite{winter:ID} however, a quite startling
discovery was made: it was shown that --- contrary to the impression
the earlier papers\cite{Loeber,AW} gave --- the identification capacity
of a (discrete memoryless, as always in this paper) quantum
channel is in general \emph{not equal to its transmission capacity}.
Indeed, the identification capacity of a noiseless qubit was found to
be $2$. This means that for quantum channels the rule that
identification capacity equals common randomness capacity
(see the discussion by Ahlswede\cite{ahlswede:GIT}
and Kleinew\"achter\cite{Kleinewaechter})
fails dramatically, even for the most ordinary of channels!
\par
In the present paper we find some new results for identification
via quantum systems: after a review of the necessary definitions and
known results (section~\ref{sec:review}) and a collection of
statements about what we called ``random channels'' in our ealier
paper,\cite{winter:ID} we first give a direct
proof that a qubit has identification capacity $2$, in
section~\ref{sec:qubit}. (Our earlier proof\cite{winter:ID} uses
a reduction to \emph{quantum identification}, which we avoid here.)
Then, in section~\ref{sec:ebit}, we show the quantum analogue of
Ahlswede and Dueck's result that $1$ bit of shared randomness
plus negligible communication are sufficient to build an identification
code of rate $1$:\cite{AD2}
namely, $1$ bit of entanglement plus negligible (quantum)
communication are sufficient to build an identification code
of rate $2$.
In section~\ref{sec:prior} we briefly discuss the case of more general
prior correlations between sender and receiver.
\par
In section~\ref{sec:feedback:qc}, we turn our attention to
feedback channels: we first study quantum--classical channels
with passive classical feedback, and prove a quantum generalisation
of the capacity formula of Ahlswede and Dueck\cite{AD2}.
Then, in section~\ref{sec:feedback:general}, we introduce
a feedback model for general quantum channels which we call
``coherent feedback'', and prove a capacity formula
for these channels as well which can be understood as a quantum
analogue of the feedback identification capacity of Ahlswede
and Dueck.\cite{AD2} We also comment on a different feedback
model for classical--quantum channels.

\section{Review of definitions and known facts}
\label{sec:review}
For a broader review of identification (and, for comparison,
transmission) via quantum channels we refer the reader to the introductory sections
of our earlier paper,\cite{winter:ID} to L\"ober's Ph.D.~thesis,\cite{Loeber}
and to the classical identification papers by Ahlswede and Dueck.\cite{AD1,AD2}
Here we are content with repeating the bare definitions:
\par
We are concerned with quantum systems, which are modelled as (finite)
Hilbert spaces ${\cal H}$ (or rather the operator algebra
${\cal B}({\cal H})$). \emph{States} on these systems we
identify with density operators $\rho$:
positive semidefinite operators with trace $1$.
\par
A \emph{quantum channel} is modelled in this context as a completely
postive, trace preserving linear map
$T:{\cal B}({\cal H}_1)\longrightarrow{\cal B}({\cal H}_2)$
between the operator algebras of Hilbert spaces ${\cal H}_1$,
${\cal H}_2$.
\begin{defi}[L\"ober,\cite{Loeber} Ahlswede and Winter\cite{AW}]
  \label{defi:IDcode}
  An \emph{identification code for the channel}
  $T$ with \emph{error probability $\lambda_1$
  of first, and $\lambda_2$ of second kind} is a set
  $\{(\rho_i,D_i):i=1,\ldots,N\}$ of states $\rho_i$ on ${\cal H}_1$ and
  operators $D_i$ on ${\cal H}_2$ with $0\leq D_i\leq \1$, such that
  \begin{align*}
    \forall i       &\quad \tr\bigl( T(\rho_i) D_i \bigr) \geq 1-\lambda_1, \\
    \forall i\neq j &\quad \tr\bigl( T(\rho_i) D_j \bigr) \leq \lambda_2.
  \end{align*}
  For the identity channel $\id_{\C^d}$ of the algebra ${\cal B}(\C^d)$
  of a $d$--dimensional system
  we also speak of an \emph{identification code on $\C^d$}.
  \par
  For the special case of memoryless channels $T^{\otimes n}$
  (where $T$ is implicitly fixed), we speak of an
  \emph{$(n,\lambda_1,\lambda_2)$--ID code}, and denote
  the largest size $N$ of such a code $N(n,\lambda_1,\lambda_2)$.
  \par
  An identification code as above is called \emph{simultaneous} if
  all the $D_i$ are coexistent: this means that there exists a positive
  operator valued measure (POVM) $(E_k)_{k=1}^K$ and sets
  ${\cal D}_i\subset\{1,\ldots,K\}$ such that $D_i=\sum_{k\in{\cal D}_i} E_k$.
  The largest size of a simultaneous $(n,\lambda_1,\lambda_2)$--ID code
  is denoted $N_{\rm sim}(n,\lambda_1,\lambda_2)$.
\end{defi}
\medskip
Most of the current knowledge about these concepts is summarised in
the two following theorems.
\begin{thm}[L\"ober,\cite{Loeber} Ahlswede and Winter\cite{AW}]
  \label{thm:simID}
  Consider any channel $T$, with transmission capacity $C(T)$
  (Holevo,\cite{Holevo98} Schumacher and Westmoreland\cite{SW97}).
  Then, the \emph{simultaneous identification capacity} of $T$,
  $$C_{\rm sim-ID}(T):=\inf_{\lambda_1,\lambda_2>0}
                       \liminf_{n\rightarrow\infty}
                       \frac{1}{n} \log\log N_{\rm sim}(n,\lambda_1,\lambda_2)
    \geq C(T).$$
  (With $\log$ and $\exp$ in this paper understood to basis $2$.)
  \par
  For classical--quantum (cq) channels $T$
  (see Holevo\cite{Holevo77}),
  even the strong converse for (non--simultaneous) identification holds:
  $$C_{\rm ID}(T)=
    \lim_{n\rightarrow\infty} \frac{1}{n} \log\log N(n,\lambda_1,\lambda_2) = C(T),$$
  whenever $\lambda_1,\lambda_2>0$ and $\lambda_1+\lambda_2<1$.
  \qed
\end{thm}
\medskip
That the (non--simultaneous) identification capacity can be
larger than the transmission capacity was shown only recently:
\begin{thm}[Winter\cite{winter:ID}]
  \label{thm:2}
  The identification capacity of the noiseless qubit channel,
  $\id_{\C^2}$, is $C_{\rm ID}(\id_{\C^2})=2$, and the strong
  converse holds.
  \qed
\end{thm}
\medskip
The main objective of the following three sections is to give two
new proofs of the achievability of $2$ in this theorem.

\section{Random channels and auxiliary results}
\label{sec:rand-channel}
The main tool in the following results (as in our earlier paper\cite{winter:ID}) are
\emph{random channels} and in fact
\emph{random states}:\cite{random:states,quantum-SD}
\begin{defi}
  \label{defi:random:channel}
  For positive integers $s,t,u$ with $s\leq tu$, the \emph{random channel}
  $R_s^{t(u)}$ is a random variable taking values in quantum channels
  ${\cal B}(\C^s)\longrightarrow{\cal B}(\C^t)$ with the following
  distribution:
  \par
  There is a random isometry $V:\C^s\longrightarrow\C^t\otimes\C^u$,
  by which we mean a random variable taking values in isometries
  whose distribution is left--/right--invariant under multiplication
  by unitaries on $\C^t\otimes\C^u$/on $\C^s$, respectively, such that
  $$R_s^{t(u)}(\rho) = \tr_{\C^u} \bigl( V\rho V^* \bigr).$$
  Note that the invariance demanded of the distribution of $V$ determines
  it uniquely --- one way to generate the distribution is to pick
  an arbitrary fixed isometry $V_0:\C^s\longrightarrow\C^t\otimes\C^u$
  and a random unitary $U$ on $\C^t\otimes\C^u$ according to the
  Haar measure, and let $V=U V_0$.
\end{defi}
\begin{rem}
  \label{rem:random:trivia:1}
  {\rm
    Identifying $\C^{tu}$ with $\C^t\otimes\C^u$, we have
    $R_s^{t(u)} = \tr_{\C^u} \circ R_s^{tu(1)}$.
    Note that $R_s^{t(1)}$ is a random isometry from $\C^s$ into
    $\C^t$ in the sense of our definition, and that the distribution
    of $R_s^{s(1)}$ is the Haar measure on the unitary group
    of $\C^s$.
  }
\end{rem}
\begin{rem}
  \label{rem:random:trivia:2}
  {\rm
    The one--dimensional Hilbert space $\C$ is a trivial
    system: it has only one state, $1$, and so the random channel
    $R_1^{t(u)}$ is equivalently described by the image state it assigns
    to $1$, $R_1^{t(u)}(1)$. For $s=1$ we shall thus identify the random channel
    $R_1^{t(u)}$ with the random state $R_1^{t(u)}(1)$ on $\C^t$.
    A different way of describing this state is that there exists a random
    (Haar distributed) unitary $U$ and a pure state $\psi_0$ such that
    $R_1^{t(u)}=\tr_{\C^u}\bigl( U \psi_0 U^* \bigr)$ ---
    note that it has rank bounded by $u$. These are the objects we
    concentrate on in the following.
  }
\end{rem}
\begin{lemma}[see Bennett \emph{et al.},\cite{rand} Winter\cite{winter:ID}]
  \label{lemma:LD}
  Let $\psi$ be a pure state, $P$ a projector of rank (at most) $r$
  and let $U$ be a random unitary,
  distributed according to the Haar measure.
  Then for $\epsilon>0$,
  \begin{equation*}
    \Pr\left\{ \tr(U \psi U^* P) \geq (1+\epsilon)\frac{r}{d} \right\}
                            \leq \exp\left( -r \frac{\epsilon-\ln(1+\epsilon)}{\ln 2}\right).
  \end{equation*}
  For $0<\epsilon\leq 1$, and $\rank P=r$,
  \begin{align*}
    \Pr\left\{ \tr(U \psi U^* P) \geq (1+\epsilon)\frac{r}{d} \right\}
                           &\leq \exp\left( -r \frac{\epsilon^2}{6\ln 2} \right), \\
    \Pr\left\{ \tr(U \psi U^* P) \leq (1-\epsilon)\frac{r}{d} \right\}
                           &\leq \exp\left( -r \frac{\epsilon^2}{6\ln 2} \right).
  \end{align*}
  \qed
\end{lemma}
\begin{lemma}[Bennett \emph{et al.}\cite{rand}]
  \label{lemma:net}
  For $\epsilon>0$, there exists in the set of pure states on $\C^d$
  an \emph{$\epsilon$--net} ${\cal M}$ of cardinality
  $|{\cal M}|\leq \left(\frac{5}{\epsilon}\right)^{2d}$; i.e.,
  $$\forall\varphi\text{ pure }\exists{\widehat\varphi\in{\cal M}}\quad
                           \| \varphi - \widehat\varphi \|_1 \leq \epsilon.$$
  \qed
\end{lemma}
With these lemmas, we can prove an important auxiliary result:
\begin{lemma}[see Harrow \emph{et al.}\cite{quantum-SD}]
  \label{lemma:uniform}
  For $0< \eta\leq 1$ and $t\leq u$,
  consider the random state $R_1^{t(u)}$ on $\C^t$. Then,
  \begin{equation*}
    \Pr\left\{ R_1^{t(u)} \not\in \left[ \frac{1-\eta}{t}\1;\frac{1+\eta}{t}\1 \right] \right\}
                     \leq 2\left(\frac{10t}{\eta}\right)^{2t}
                            \exp\left( -u \frac{\eta^2}{24\ln 2} \right).
  \end{equation*}
\end{lemma}
\begin{beweis}
  We begin with the observation that $R_1^{t(u)} \in [ \alpha\1;\beta\1 ]$ if and
  only if for all pure states (rank one projectors) $\varphi$,
  \begin{equation*}
    \tr\bigl( R_1^{t(u)}\varphi \bigr)
       = \tr\bigl( R_1^{tu(1)}(\varphi\otimes\1_u) \bigr)
          \begin{cases}
            \geq \alpha, & \\
            \leq \beta.  & 
          \end{cases}
  \end{equation*}
  Due to the triangle inequality, we have to ensure this only
  for $\varphi$ from an $\eta/2t$--net and with
  $\alpha=\left(1-\frac{\eta}{2}\right)/t$,
  $\beta=\left(1+\frac{\eta}{2}\right)/t$.
  Then the probability bound claimed above follows from
  lemmas~\ref{lemma:LD} and~\ref{lemma:net}, with the union bound.
\end{beweis}

\section{ID capacity of a qubit}
\label{sec:qubit}
Here we give a new, direct proof of theorem~\ref{thm:2} --- in fact,
we prove the following proposition from which it follows directly.
\begin{prop}
  \label{prop:2}
  For every $0<\lambda<1$, there exists on the quantum system
  ${\cal B}(\C^d)$ an ID code with
  $$N = \left\lceil \frac{1}{2}
              \exp\left( \left(\frac{\lambda}{3000}\frac{d}{\log d}\right)^2 \right)
        \right\rceil$$
  messages, with error probability of first kind equal to $0$ and error
  probability of second kind bounded by $\lambda$.
\end{prop}
\begin{beweis}
  We shall prove even a bit more: that such a code exists which is
  of the form $\{(\rho_i,D_i):i=1,\ldots,N\}$ with
  \begin{equation}
    \label{eq:properties}
    D_i = \supp \rho_i, \quad
    \rank \rho_i = \delta:=\alpha\frac{d}{\log d}, \quad
    \rho_i \leq \frac{1+\eta}{\delta}D_i.
  \end{equation}
  The constants $\alpha\leq \lambda/4$ and $\eta\leq 1/3$
  will be fixed in the course of this proof.
  Let a \emph{maximal} code ${\cal C}$ of this form be given.
  We shall show that if $N$ is ``not large'', a random codestate as follows
  will give a larger code, contradicting maximality.
  \par
  Let $R=R_1^{d(\delta)}$ (the random state in dimension $d$ with
  $\delta$--dimensional ancillary system, see definition~\ref{defi:random:channel}),
  and $D:=\supp R$. Then, according to the Schmidt decomposition
  and lemma~\ref{lemma:uniform},
  \begin{equation}\begin{split}
    \label{eq:R-evenness}
    \Pr\left\{ R\not\in\left[ \frac{1-\eta}{\delta}D; \frac{1+\eta}{\delta}D \right] \right\}
      &=    \Pr\left\{ R_1^{\delta(d)}\not\in
         \left[ \frac{1-\eta}{\delta}\1_\delta; \frac{1+\eta}{\delta}\1_\delta \right] \right\} \\
      &\leq 2\left(\frac{10\delta}{\eta}\right)^{2\delta}
               \exp\left( -d \frac{\eta^2}{24\ln 2} \right).
  \end{split}\end{equation}
  This is $\leq 1/2$ if
  $$d \geq \left(\frac{96\ln 2}{\eta^2}\log\frac{10}{\eta}\right) \delta\log\delta,$$
  which we ensure by choosing
  $\alpha \leq \lambda \left(\frac{96\ln 2}{\eta^2}\log\frac{10}{\eta}\right)^{-1}
          \leq \lambda/4$.
  \par
  In the event that $\frac{1-\eta}{\delta}D\leq R\leq \frac{1+\eta}{\delta}D$,
  we have on the one hand
  \begin{equation}
    \label{eq:rho-i:D}
    \tr(\rho_i D) \leq \tr\left( \frac{1+\eta}{\delta}D_i\frac{\delta}{1-\eta}R \right)
                 \leq 2 \tr(R D_i).
  \end{equation}
  On the other hand, because of $R_1^{d(\delta)}=\tr_{\C^\delta} R_1^{d\delta(1)}$,
  we can rewrite $\tr(R D_i) = \tr\bigl( R_1^{d\delta(1)}(D_i\otimes\1_\delta) \bigr)$,
  hence by lemma~\ref{lemma:LD}
  \begin{equation}
    \label{eq:R:D-i}
    \Pr\bigl\{ \tr(R D_i) > \lambda/2 \bigr\} \leq \exp\bigl( -\delta^2 \bigr).
  \end{equation}
  So, by the union bound, eqs.~(\ref{eq:rho-i:D}) and~(\ref{eq:R:D-i}) yield
  \begin{equation*}\begin{split}
    \Pr&\Bigl\{ {\cal C}\cup\{(R,D)\} \text{ has error probability of} \Bigr.        \\
       &\phantom{=}
        \Bigl. \text{second kind larger than }\lambda
               \text{ or violates eq. }(\ref{eq:properties}) \Bigr\}
                                       \leq \frac{1}{2} + N\exp\bigl( -\delta^2 \bigr).
  \end{split}\end{equation*}
  If this is less than $1$, there must exist a pair $(R,D)$ extending
  our code while preserving the error probabilities and the
  properties of eq.~(\ref{eq:properties}),
  which would contradict maximality.
  Hence,
  $$N\geq \frac{1}{2}\exp\left( \delta^2 \right),$$
  and we are done, fixing $\eta=1/3$ and $\alpha=\lambda/3000$.
\end{beweis}
\medskip\noindent
The \emph{proof of theorem~\ref{thm:2}} is now obtained by
applying the above proposition to $d=2^n$, the Hilbert space
dimension of $n$ qubits, and arbitarily small $\lambda$.
That the capacity is not more than $2$ is by a simple dimension counting
argument,\cite{winter:ID} which we don't repeat here.
\qed

\section{ID capacity of an ebit}
\label{sec:ebit}
Ahlswede and Dueck\cite{AD2} have shown that the identification capacity
of any system, as soon as it allows --- even negligible --- communication,
is at least as large as its common randomness capacity: the maximum rate
at which shared randomness can be generated. (We may add, that except
for pathological examples expressly constructed for that purpose,
in practically all classical systems for which these two capacities exist, they
turn out to be equal.\cite{AD2,ahlswede:balakirsky,Kleinewaechter,ahlswede:GIT})
\par
Their proof relies on a rather general construction, which we restate here,
in a simplified version:
\begin{prop}[Ahlswede and Dueck\cite{AD2}]
  \label{prop:AD2}
  There exist, for $\lambda>0$ and $N\geq 4^{1/\lambda}$, functions
  $f_i:\{1,\ldots,M\}\longrightarrow\{1,\ldots,N\}$ ($i=1,\ldots,2^M$)
  such that the distributions $P_i$ on $\{1,\ldots,M\}\times\{1,\ldots,N\}$
  defined by
  \begin{equation*}
    P_i(\mu,\nu) = \begin{cases}
                     \frac{1}{M} & \text{ if } \nu=f_i(\mu), \\
                     0           & \text{ otherwise}.
                   \end{cases}
  \end{equation*}
  and the sets ${\cal D}_i=\supp P_i$ form an identification code
  with error probability of first kind $0$ and error probability
  of second kind $\lambda$.
  \par
  In other words, prior shared randomness in the form of uniformly distributed
  $\mu\in\{1,\ldots,M\}$ between sender and receiver, and transmission of
  $\nu\in\{1,\ldots,N\}$ allow identification of $2^M$ messages.
  \qed
\end{prop}
(In the above form it follows from proposition~\ref{prop:blowup-code}
below: a perfect transmission code is at the same time always an
identification code with both error probabilities $0$.)
\par
Thus, an alternative way to prove that a channel of capacity $C$ allows
identification at rate $\geq C$, is given by the following scheme:
use the channel $n-O(1)$ times to generate $Cn-o(n)$ shared random bits
and the remaining $O(1)$ times to transmit one out of $N=2^{O(1)}$ messages;
then apply the above construction with $M=2^{Cn-o(n)}$. More generally,
a rate $R$ of common randomness and only negligible communication give
identification codes of rate $R$.
\par
The quantum analogue of perfect correlation (i.e., shared randomness) being
pure entanglement, substituting quantum state transmission wherever
classical information was conveyed, and in the light of the result that a
qubit has identification capacity $2$, the following question appears rather
natural (and we have indeed raised it, in remark 14 of our earlier
paper\cite{winter:ID}):
Does $1$ bit of entanglement plus the ability to (even only negligibly)
communicate result in an ID code of rate $2$, asymptotically?
\begin{prop}
  \label{prop:quantum-hashing}
  For $\lambda>0$, $d\geq 2$ and
  $\Delta \geq \left(\frac{900}{\lambda^2}\log\frac{30d}{\lambda}\right) \log d$,
  there exist quantum channels $T_i:{\cal B}(\C^d)\longrightarrow{\cal B}(\C^\Delta)$
  ($i=1,\ldots,N'=\left\lceil \frac{1}{2}\exp(d^2) \right\rceil$),
  such that the states $\rho_i = (\id \otimes T_i)\Phi_d$
  (with state vector $\ket{\Phi_d} = \frac{1}{\sqrt{d}}\sum_{j=1}^d \ket{j}\ket{j}$), and
  the operators $D_i=\supp\rho_i$ form an identification
  code on ${\cal B}(\C^d\otimes\C^\Delta)$ with
  error probability of first kind $0$ and error probability
  of second kind $\lambda$.
  \par
  In other words, sender and receiver, initially sharing the maximally entangled
  state $\Phi_d$, can use transmission of a $\Delta$-dimensional
  system to build an identification code with
  $\left\lceil \frac{1}{2}\exp(d^2) \right\rceil$ messages.
\end{prop}
\begin{beweis}
  Let a maximal code ${\cal C}$ as described in the proposition be given,
  such that additionally
  \begin{equation}
    \label{eq:properties:e}
    D_i = \supp \rho_i, \quad
    \rank \rho_i = d, \quad
    \rho_i \leq \frac{1+\lambda}{d}D_i.
  \end{equation}
  Consider the random state $R=R_1^{d\Delta(d)}$ on
  $\C^{d\Delta}=\C^d\otimes\C^\Delta$, and $D:=\supp R$.
  Now, by Schmidt decomposition and with lemma~\ref{lemma:LD}
  (compare the proof of proposition~\ref{prop:2}),
  for $\eta:=\lambda/3$
  \begin{equation}\begin{split}
    \label{eq:R-evenness:e}
    \Pr\left\{ R\not\in\left[ \frac{1-\eta}{d}D; \frac{1+\eta}{d}D \right] \right\}
      &=    \Pr\left\{ R_1^{d(\Delta d)}\not\in
         \left[ \frac{1-\eta}{d}\1_d; \frac{1+d}{\delta}\1_d \right] \right\} \\
      &\leq 2\left(\frac{10d}{\eta}\right)^{2d}
               \exp\left( -d\Delta \frac{\eta^2}{24\ln 2} \right).
  \end{split}\end{equation}
  The very same estimate gives
  \begin{equation}\begin{split}
    \label{eq:R-evenness:ee}
    \Pr\!\left\{\! \tr_{\C^\Delta} R \!\not\in\!
                 \left[ \frac{1-\eta}{d}\1_d; \frac{1+\eta}{d}\1_d \right] \right\}
      &=    \Pr\!\left\{\! R_1^{d(\Delta d)} \!\not\in\!
                 \left[ \frac{1-\eta}{d}\1_d; \frac{1+\eta}{d}\1_d \right] \right\} \\
      &\leq 2\left(\frac{10d}{\eta}\right)^{2d}
               \exp\left( -d\Delta \frac{\eta^2}{24\ln 2} \right).
  \end{split}\end{equation}
  By choosing $\Delta \geq \left(\frac{144\ln 2}{\eta^2}\log\frac{10}{\eta}\right) \log d$,
  as we indeed did, the sum of these two probabilities is at most $1/2$.
  \par
  In the event that $\frac{1-\eta}{d}D \leq R \leq \frac{1+\eta}{d}D$, we
  argue similar to the proof of proposition~\ref{prop:2}
  (compare eq.~(\ref{eq:rho-i:D})):
  \begin{equation}
    \label{eq:rho-i:D:e}
    \tr(\rho_i D) \leq \tr\left( \frac{1+\lambda}{d}D_i \frac{d}{1-\eta}R\right)
                  \leq 3\tr(R D_i).
  \end{equation}
  On the other hand (compare eq.~(\ref{eq:R:D-i})),
  \begin{equation}
    \label{eq:R:D-i:e}
    \Pr\bigl\{ \tr(R D_i) > \lambda/3 \bigr\} \leq \exp\bigl( -d^2 \bigr),
  \end{equation}
  by lemma~\ref{lemma:LD} and using $\Delta^{-1}\leq \lambda/6$.
  \par
  In the event that
  $\frac{1-\eta}{d}\1 \leq \tr_{\C^\Delta} R \leq \frac{1+\eta}{d}\1$,
  there exists an operator $X$ on $\C^d$ with
  $\frac{1}{1+\eta}\1 \leq X \leq \frac{1}{1-\eta}\1$, such that
  $$R_0 := \sqrt{R} (X\otimes\1) \sqrt{R}\ \,(\text{which has the same support }D\text{ as }R)$$
  satisfies $\tr_{\C^\Delta} R_0 = \frac{1}{d}\1$. By the
  Jamio\l{}kowski isomorphism\cite{jamiolkowski} between quantum channels and
  states with maximally mixed reduction, this is equivalent to the existence 
  of a quantum channel $T_0$ such that $R_0=(\id\otimes T_0)\Phi_d$.
  Observe that $R_0\leq \frac{1+\lambda}{d}D$ and
  $\tr(R_0 D_i)\leq \frac{3}{2}\tr(R D_i)$.
  \par
  So, putting together the bounds of eqs.~(\ref{eq:R-evenness:e}),
  (\ref{eq:R-evenness:ee}),
  (\ref{eq:rho-i:D:e}) and~(\ref{eq:R:D-i:e}), we get, by the union bound,
  \begin{equation*}\begin{split}
    \Pr&\Bigl\{ {\cal C}\cup\{(R_0,D)\} \text{ has error probability of} \Bigr. \\
       &\phantom{=}
        \Bigl. \text{second kind larger than }\lambda
               \text{ or violates eq. }(\ref{eq:properties:e})\Bigr\}
                             \leq \frac{1}{2} + N'\exp\bigl( -d^2 \bigr).
  \end{split}\end{equation*}
  If this is less than $1$, there will exist a state $R_0=(\id\otimes T_0)\Phi_d$
  and an operator $D$ enlarging the code and preserving the error probabilities
  as well as the properties
  in eq.~(\ref{eq:properties:e}), which contradicts maximality.
  \par
  Hence, $N'\geq \frac{1}{2}\exp\bigl(d^2\bigr)$, and we are done.
\end{beweis}
This readily proves, answering the above question affirmatively:
\begin{thm}
  \label{thm:2-e}
  The identification capacity of a system in which entanglement (EPR pairs) between
  sender and receiver is available at rate $E$, and which allows (even only
  negligible) communication, is at least $2E$.
  This is tight for the case that the available resources are only
  the entanglement and negligible communication.
  \qed
\end{thm}
\begin{rem}
  \label{rem:hashing}
  {\rm
  Just as the Ahlswede--Dueck construction of
  proposition~\ref{prop:AD2} can be understood as
  an application of random hashing, we are tempted to present
  our above construction as a kind of ``quantum hashing'': indeed,
  the (small) quantum system transmitted contains, when held
  together with the other half of the prior shared entanglement,
  just enough of a signature of the functions/quantum channels
  used to distinguish them pairwise reliably.
  }
\end{rem}

\section{General prior correlation}
\label{sec:prior}
Proposition~\ref{prop:AD2} quantifies the identification capacity of
shared randomness, and proposition~\ref{prop:quantum-hashing}
does the same for shared (pure) entanglement.
This of course raises the questions what the identification capacity
of other, more general, correlations is: i.e., we are asking for
code constructions and bounds if (negligible) quantum communication
and $n$ copies of a bipartite state $\omega$ between sender and receiver are
available.
\par
For the special case that the correlation decomposes cleanly into
entanglement and shared randomness,
$$\omega=\sum_{\mu} p_\mu \Psi_\mu^{AB} \otimes \ketbra{\mu}^{A'}\otimes\ketbra{\mu}^{B'},$$
with an arbitrary perfect classical correlation (between registers $A'$ and $B'$)
distributed according to $p$ and arbitrary pure entangled states
$\Psi_\mu$, we can easily give the answer (let the sender be in possession of
$AA'$, the receiver of $BB'$):
\begin{equation}
  \label{eq:E-and-R}
  C_{\rm ID} = H(p) + 2 \sum_\mu p_\mu E(\Psi_\mu^{AB});
\end{equation}
here, $H(p)$ is the entropy of the classical perfect correlation $p$;
$E(\Psi^{AB})=S(\Psi^A)$ is the entropy of entanglement,\cite{concentration}
with the reduced state $\Psi^A=\tr_B \Psi^{AB}$.
The achievability is seen as follows: by entanglement and randomness
concentration\cite{concentration} this state yields shared
randomness and entanglement at rates $R=H(p)$ and $E=\sum_\mu p_\mu E(\Psi_\mu)$,
respectively (without the need of communication --- note that both
users learn which entangled state they have by looking at the
primed registers).
Proposition~\ref{prop:quantum-hashing} yields an identification code
of rate $2E$, while proposition~\ref{prop:blowup-code} below shows
how to increase this rate by $R$.
\par
That the expression is an upper bound is then easy to see, along the lines
of the arguments given in our earlier paper for the capacity of
a ``hybrid quantum memory''.\cite{kuperberg,winter:ID}
\begin{prop}[Winter\cite{winter:ID}]
  \label{prop:blowup-code}
  Let $\{(\rho_i,D_i):i=1,\ldots,N\}$
  be an identification code on the quantum system ${\cal H}$ with error probabilities
  $\lambda_1,\lambda_2$ of first and second kind, respectively,
  and let ${\cal H}_C$ be a classical system of dimension $M$
  (by this we mean a Hibert space only allowed to be in a state
  from a distinguished orthonormal basis $\{\ket{\mu}\}_{\mu=1}^M$).
  Then, for every $\epsilon>0$, there exists an identification code
  $\{(\sigma_f,\widetilde{D}_f):f=1,\ldots,N'\}$
  on ${\cal H}_C\otimes{\cal H}$ with error probabilities
  $\lambda_1,\lambda_2+\epsilon$ of first and second kind, respectively,
  and $N'\geq\left(\frac{1}{2}N^\epsilon\right)^M$.
  The $f$ actually label functions (also denoted $f$)
  $\{1,\ldots,M\}\longrightarrow\{1,\ldots,N\}$, such that
  $$\sigma_f = \frac{1}{M} \sum_\mu \ketbra{\mu} \otimes \rho_{f(k)}.$$
  \par
  In other words, availability of shared randomness ($\mu$ on the classical
  system ${\cal H}_C$) with an identification
  code allows us to construct a larger identification code.
  \qed
\end{prop}
\medskip
The general case seems to be much more complex, and we cannot offer
an approximation to the solution here. So, we restrict ourselves
to highlighting two questions for further investigation:
\begin{enumerate}
 \item What is the identification capacity of a bipartite state $\omega$,
   together with negligible communication? For noisy correlations,
   this may not be the right question altogether, as a look at work by
   Ahlswede and Balakirsky\cite{ahlswede:balakirsky} shows: they have studied this
   problem for classical binary correlations with symmetric noise, and have found
   that --- as in common randomness theory\cite{AC} --- one ought to
   include a limited rate of communication and study the relation between
   this additional rate and the obtained identification rate.
   Hence, we should ask: what is the identification capacity of $\omega$
   plus a rate of $C$ bits of communication?
   An obvious thing to do in this scenario would be to use part of this rate
   to do entanglement distillation of which the communication
   cost is known in principle.\cite{hashing} This gives entanglement as well
   as shared randomness, so one can use the constructions above.
   It is not clear of course whether this is asymptotically optimal.
  \item In the light of the code enlargement proposition~\ref{prop:blowup-code},
   it would be most interesting to know if a stronger version of our
   proposition~\ref{prop:quantum-hashing}/theorem~\ref{thm:2-e} holds:
   Does entanglement of rate $E$ increase the rate of a given identification
   code by $2E$?
\end{enumerate}

\section{Identification in the presence of feedback:\protect\\
         quantum--classical channels}
\label{sec:feedback:qc}
Feedback for quantum channels is a somewhat problematic issue,
mainly because the output of the channel is a quantum state,
of which there is in general no physically consistent way of giving
a copy to the sender. In addition, it should not even be a ``copy'' for the
general case that the channel outputs a mixed state (which corresponds to
the distribution of the output), but a copy of the
exact \emph{symbol} the receiver obtained; so the feedback should establish
correlation between sender and receiver, and in the quantum case this
appears to involve further choices, e.g.~of basis.
The approach taken in the small literature on the issue of feedback
in quantum channels (see Fujiwara
and Nagaoka,\cite{fujiwara:nagaoka} Bowen,\cite{bowen}
and Bowen and Nagarajan\cite{bowen:nagarajan})
has largely been to look at active feedback, where the receiver decides
what to give back to the sender, based on a partial evaluation of the
received data.
\par
We will begin our study by looking at a subclass of channels which
do not lead into any of these conceptual problems: quantum--classical (qc) channels,
i.e., destructive measurements, have a completely classical output anyway,
so there is no problem in augmenting every use of the channel by
instantaneous passive feedback.
\par
Let a measurement POVM $(M_y)_{y\in{\cal Y}}$ be given; then its qc--channel
is the map
$$T:\rho \longmapsto \sum_y \tr(\rho M_y) \ketbra{y},$$
with an orthogonal basis $(\ket{y})_y$ of an appropriate Hilbert space ${\cal F}$, say.
We will denote this qc--channel as $T:{\cal B}({\cal H})\longrightarrow{\cal Y}$.
\par
For a qc--channels $T$, a \emph{(randomised) feedback strategy}
$F$ for block $n$ is given by states $\rho_{t:y^{t-1}}$ on ${\cal H}_1$
for each $t=1,\ldots,n$ and $y^{t-1}\in{\cal Y}^{t-1}$: this is the
state input to the channel in the $t^{\rm th}$ timestep if the
feedback from the previous rounds was $y^{t-1}=y_1\ldots y_{t-1}$.
Clearly, this defines an output distribution $Q$ on ${\cal Y}^n$
by iteration of the feedback loop:
\begin{equation}
  \label{eq:feedback-out}
  Q(y^n) = \prod_{t=1}^n \tr\bigl( \rho_{t:y^{t-1}} M_{y_t} \bigr).
\end{equation}
\begin{rem}
  \label{rem:feedback-strategy}
  {\rm
  We could imagine a more general protocol for the sender: an initial
  state $\sigma_0$ could be prepared on an ancillary system ${\cal H}_A$,
  and the feedback strategy is a collection $\Phi$ of completely positive, trace
  preserving maps
  $$\varphi_t:{\cal B}\bigl({\cal F}^{\otimes (t-1)} \otimes {\cal H}_A\bigr)
                        \longrightarrow {\cal B}\bigl( {\cal H}_A \otimes {\cal H} \bigr),$$
  where ${\cal F}$ is the quantum system representing the classical feedback
  by states from an orthogonal basis: this map creates the next channel input
  and a new state of the ancilla (potentially entangled)
  from the old ancilla state and the feedback.
  \par
  This more general scheme allows for memory and even quantum correlations
  between successive uses of the channel, via the system ${\cal H}_A$.
  However, the scheme has, for each ``feedback history'' $y^{t-1}$ up to time $t$,
  a certain state $\sigma_{t-1:y^{t-1}}$ on ${\cal H}_A$ (starting with $\sigma_0$),
  and consequently an input state $\rho_{t:y^{t-1}}$ on ${\cal H}$:
  \begin{align*}
    \rho_{t:y^{t-1}}  &= \tr_{{\cal H}_A}\Bigl( \varphi_t\bigl(
                                                               \ketbra{y^{t-1}} \otimes \sigma_{t-1:y^{t-1}} \bigr) \Bigr), \\
    \sigma_{t:y^{t}} &= \frac{1}{\tr\bigl( \rho_{t:y^{t-1}} M_{y_t} \bigr)}
                                          \tr_{{\cal H}}\Bigl( \varphi_t\bigl(
                                                               \ketbra{y^{t-1}} \otimes \sigma_{t-1:y^{t-1}} \bigr) \Bigr).
  \end{align*}
  It is easy to check that the corresponding output distribution $Q$
  of this feedback strategy according to our definition (see eq.~(\ref{eq:feedback-out}))
  is the same as for the original, more general feedback scheme. So, we
  do not need to consider those to obtain ultimate generality.
  }
\end{rem}
\par\medskip
An \emph{$(n,\lambda_1,\lambda_2)$--feedback ID code} for
the qc--channel $T$ with passive feedback is now a set
$\{(F_i,D_i):i=1,\ldots,N\}$ of feedback strategies $F_i$ and
of operators $0 \leq D_i  \leq \1$,
such that the output states $\omega_i = \sum_{y^n} Q_i(y^n)\ketbra{y^n}$
with the operators $D_i$ form an identification code with
error probabilities $\lambda_1$ and $\lambda_2$
of first and second kind.
Note that because the output is classical --- i.e., the states are
diagonal in the basis $(\ket{y^n})$ ---, we may without loss of generality
assume that all $D_i = \sum_{y^n} D_i(y^n)\ketbra{y^n}$, with certain
$0 \leq D_i(y^n) \leq 1$.
\par
Finally, let $N_F(n,\lambda_1,\lambda_2)$ be the maximal $N$ such that
there exists an $(n,\lambda_1,\lambda_2)$--feedback ID code
with $N$ messages. Note that due to the classical nature of the channel output
codes are automatically simultaneous.
\par
To determine the capacity, we invoke the following result:
\begin{lemma}[Ahlswede and Dueck,\cite{AD2} Lemma 4]
  \label{lemma:product-qc}
  Consider a qc--channel $T:{\cal B}({\cal H})\rightarrow {\cal Y}$
  and any randomised feedback strategy $F$ for block $n$.
  Then, for $\epsilon>0$, there exists a set ${\cal E}\subset{\cal Y}^n$ of
  probability $Q({\cal E}) \geq 1-\epsilon$ and cardinality
  $$|{\cal E}| \leq \exp\left( n \max_\rho H(T(\rho))+ \alpha\sqrt{n} \right),$$
  where $\alpha = |{\cal Y}| \epsilon^{-1/2}$.
\end{lemma}
  The \emph{proof} of Ahlswede and Dueck\cite{AD2} applies directly:
  a qc--channel with feedback is isomorphic to a classical feedback channel with an
  infinite input alphabet (the set of all states), but with finite output alphabet,
  which is the relevant fact. \qed
\par\medskip\noindent
This is the essential tool to prove the following generalisation of
Ahlswede's and Dueck's capacity result:\cite{AD2}
\begin{thm}
  \label{thm:qc-feedback}
  For a qc--channel $T$ and $\lambda_1,\lambda_2>0$, $\lambda_1+\lambda_2<1$,
  $$\lim_{n\rightarrow\infty} \frac{1}{n}\log\log N_F(n,\lambda_1,\lambda_2)
                                 = C_{\rm ID}^F(T) = \max_\rho H\bigl(T(\rho)\bigr),$$
  unless the transmission capacity of $T$ is $0$, in which case
  $C_{\rm ID}^F(T)=0$.
  \par
  In other words, the capacity of a nontrivial qc--channel with feedback is
  its maximum output entropy and the strong converse holds.
\end{thm}
\begin{beweis}
  Let's first get the exceptional case out of the way: $C(T)$
  can only be $0$ for a constant channel (i.e., one mapping every input to
  the same output). Clearly such a channel allows not only no transmission
  but also no identification.
  \par
  The achievability is explained in the paper of Ahlswede and Dueck:\cite{AD2}
  the sender uses $m=n-O(1)$ instances of the channel with the state
  $\rho$ each, which maximises the output entropy. Due to feedback they
  then share the outcomes of $m$ i.i.d.~random experiments, which they
  can concentrate into $n H(T(\rho)) - o(n)$ uniformly distributed bits.
  (This is a bit simpler than in the original paper:\cite{AD2} they just
  cut up the space into type classes.)
  The remaining $O(1)$ uses of the channel
  (with an appropriate error correcting code)
  are then used to implement
  the identification code of proposition~\ref{prop:AD2} based on the
  uniform shared randomness.
  \par
  The strong converse is only a slight modification of the arguments of Ahlswede and
  Dueck,\cite{AD2} due to the fact that we allow probabilistic decoding
  procedures: first, for each message $i$ in a given code,
  lemma~\ref{lemma:product-qc} gives us a set ${\cal E}_i\subset{\cal Y}^n$
  of cardinality
  $\leq K=\exp\left( n\max_\rho H(T(\rho)) + 3|{\cal Y}|\epsilon^{-1/2}\sqrt{n} \right)$,
  with probability $1-\epsilon/3$ under the feedback strategy $F_i$,
  where $\epsilon:=1-\lambda_1-\lambda_2>0$. Now let
  $c:=\lceil \frac{3}{\epsilon} \rceil$, and define new decoding
  rules by letting
  \begin{equation*}
    \widehat{D}_i(y^n) := \begin{cases}
                            \frac{1}{c}\lfloor c D_i(y^n) \rfloor
                                          &\text{ for }y^n\in{\cal E}_i,    \\
                            0             &\text{ for }y^n\not\in{\cal E}_i.
                          \end{cases}
  \end{equation*}
  (I.e., round the density $D_i(y^n)$ down to the nearest
  multiple of $1/c$ within ${\cal E}_i$, and to $0$ without.)
  It is straightforward to check that in this way we obtain an
  $\bigl(n,\lambda_1+\frac{2}{3}\epsilon,\lambda_2\bigr)$--feedback ID code.
  \par
  The argument is concluded by observing that the new decoding
  densities are (i) all distinct (otherwise
  $\lambda_1+\frac{2}{3}\epsilon+\lambda_2\geq 1$), and (ii) all have support
  $\leq K=\exp\left( n\max_\rho H(T(\rho)) + 3|{\cal Y}|\epsilon^{-1/2}\sqrt{n} \right)$.
  Hence
  $$N \leq \binom{|{\cal Y}|^n}{K}(c+1)^K
      \leq \Bigl[ (c+1)|{\cal Y}|^n \Bigr]^{2^{n \max_\rho H(T(\rho)) + O(\sqrt{n})}},$$
  from which the claim follows.
\end{beweis}

\section{Identification in the presence of feedback:\protect\\
         ``coherent feedback channels''}
\label{sec:feedback:general}
Inspired by the work of Harrow\cite{ccc} we propose the following definition
of ``coherent feedback'' as a substitute for full passive feedback:
by Stinespring's theorem we can view the channel $T$ as an isometry
$U:{\cal H}_1 \longrightarrow {\cal H}_2\otimes{\cal H}_3$, followed
by the partial trace $\tr_3$ over ${\cal H}_3$:
$T(\rho) = \tr_3 \bigl( U \rho U^* \bigr)$.
\emph{Coherent feedback} is now defined as \emph{distributing}, on input $\rho$,
the bipartite state $\Theta(\rho) := U\rho U^*$ among sender and receiver, who
get ${\cal H}_3$ and ${\cal H}_2$, respectively.
\par
A \emph{coherent feedback strategy} $\Phi$ for block $n$ consists of a system ${\cal H}_A$,
initially in state $\sigma_0$, and quantum channels
$$\varphi_t: {\cal B}\bigl( {\cal H}_A\otimes{\cal H}_3^{\otimes(t-1)} \bigr)
          \longrightarrow {\cal B}\bigl( {\cal H}_A\otimes{\cal H}_3^{\otimes(t-1)}\otimes{\cal H}_1 \bigr),$$
creating the $t^{\rm th}$ round channel input from the memory in ${\cal H}_A$
and the previous coherent feedback ${\cal H}_3^{\otimes(t-1)}$.
The output state on ${\cal H}_2^{\otimes n}$ after $n$ rounds of coherent
feedback channel alternating with the $\varphi_t$, is
$$\omega = \tr_{{\cal H}_A\otimes{\cal H}_3^{\otimes n}}
                     \Bigl[ \bigl( \Theta \circ \varphi_n
                                          \circ \Theta \circ \varphi_{n-1}
                                          \circ \cdots
                                          \circ \Theta \circ \varphi_1 \bigr)\sigma_0 \Bigr],$$
where implicitly each $\Theta$ is patched up by an identity on all
systems different from ${\cal H}_1$, and each $\varphi_t$ is patched up
by an identity on ${\cal H}_2^{\otimes(t-1)}$.
\par
Now, an \emph{$(n,\lambda_1,\lambda_2)$--coherent feedback ID code} for the channel
$T$ with coherent feedback consists of $N$ pairs $(\Phi_i,D_i)$ of coherent
feedback strategies $\Phi_i$ (with output states $\omega_i$)
and operators $0 \leq D_i \leq \1$ on ${\cal H}_2^{\otimes n}$, such that
the $(\omega_i,D_i)$ form an $(n,\lambda_1,\lambda_2)$--ID code on
${\cal H}_2^{\otimes n}$.
\par
As usual, we introduce the maximum size $N$
of an $(n,\lambda_1,\lambda_2)$--coherent feedback ID code,
and denote it $N_{\ket{F}}(n,\lambda_1,\lambda_2)$.
It is important to understand the difference to $N_F(n,\lambda_1,\lambda_2)$
at this point: for the qc--channel, the latter refers to codes
making use of the classical feedback of the measurement result,
but coherent feedback --- even for qc--channels --- creates entanglement
between sender and receiver, which, as we have seen in section~\ref{sec:ebit},
allows for larger identification codes.
\par
We begin by proving the analogue of lemma~\ref{lemma:product-qc}:
\begin{lemma}
  \label{lemma:product}
  Consider a quantum channel $T:{\cal B}({\cal H}_1)\rightarrow {\cal B}({\cal H}_2)$
  and any feedback strategy $\Phi$ on block $n$ with output state
  $\omega$ on ${\cal H}_2^{\otimes n}$.
  Then, for $\epsilon>0$, there exists a projector $\Pi$ on
  ${\cal H}_2^{\otimes n}$ with
  probability $\tr(\omega\Pi) \geq 1-\epsilon$ and rank
  $$\rank\Pi \leq \exp\left( n\max_\rho S(T(\rho)) + \alpha\sqrt{n} \right),$$
  where $\alpha = (\dim{\cal H}_2)\epsilon^{-1/2}$.
\end{lemma}
\begin{beweis}
  The feedback strategy
  determines the output state $\omega$ on
  ${\cal H}_2^{\otimes n}$, and we choose complete von Neumann measurements
  on each of the $n$ tensor factors: namely,
  the measurement $M$ of an eigenbasis $(\ket{m_y})_y$
  of $\widetilde\omega$, the entropy--maximising output state of $T$
  (which is unique, as easily follows from the strict concavity of $S$).
  \par
  Defining the qc--channel $\widetilde{T} := M \circ T$ (i.e., the channel
  $T$ followed by the measurement $M$), we are in the
  situation of lemma~\ref{lemma:product-qc}, with
  ${\cal Y}=\{1,\ldots,\dim{\cal H}_2\}$.
  Indeed, we can transform the given quantum feedback strategy into
  one based solely on the classical feedback of the
  measurement results, as explained in remark~\ref{rem:feedback-strategy}.
  Note that the additional quantum information available now at the sender
  due to the coherent feedback does not impair the validity of the
  argument of that remark: the important thing is that the classical feedback
  of the measurement results collapses the sender's state into one depending
  only on the message and the feedback.
  \par
  By lemma~\ref{lemma:entropy-max} stated below,
  $\max_\rho H\bigl( \widetilde{T}(\rho) \bigr) = S(\widetilde\omega)$, so
  lemma~\ref{lemma:product-qc} gives us a set ${\cal E}$ of probability
  $Q({\cal E}) \geq 1-\epsilon$ and
  $|{\cal E}| \leq \exp\bigl( n S(\widetilde\omega) + \alpha\sqrt{n} \bigr)$.
  The operator
  $$\Pi := \sum_{y^n\in{\cal E}}
            \ketbra{m_{y_1}}\otimes\cdots\otimes\ketbra{m_{y_n}}$$
  then clearly satisfies $\tr(\omega\Pi) = Q({\cal E}) \geq 1-\epsilon$, and
  $\rank\Pi = |{\cal E}|$ is bounded as in lemma~\ref{lemma:product-qc}.
\end{beweis}
\begin{lemma}
  \label{lemma:entropy-max}
  Let $T:{\cal B}(\C^{d_1}) \longrightarrow {\cal B}(\C^{d_2}) $
  be a quantum channel and let $\widetilde\rho$ maximise
  $S(T(\rho))$ among all input states $\rho$. Denote $\widetilde\omega=T(\widetilde\rho)$
  (which is easily seen to be the unique entropy--maximising output state of $T$),
  and choose a diagonalisation $\widetilde\omega = \sum_j \lambda_j\ketbra{e_j}$.
  Then, for the channel $\widetilde{T}$ defined by
  $$\widetilde{T}(\rho) = \sum_j \ketbra{e_j}\, T(\rho) \,\ketbra{e_j}$$
  (i.e., $T$ followed by dephasing of the eigenbasis of $\widetilde{\omega}$),
  $$\max_\rho S\bigl( \widetilde{T}(\rho) \bigr)
                  = S(\widetilde\omega) = \max_\rho S\bigl( T(\rho) \bigr).$$
\end{lemma}
\begin{beweis}
  The inequality ``$\geq$'' is trivial because for input state $\widetilde\rho$,
  $T$ and $\widetilde{T}$ have the same output state.
  \par
  For the opposite inequality, let us first deal with the case that $\widetilde\omega$
  is strictly positive (i.e., $0$ is not an eigenvalue). The lemma is trivial if
  $\widetilde\omega = \frac{1}{d_2}\1$, so we assume $\widetilde\omega \neq \frac{1}{d_2}\1$
  from now on.
  Observe that ${\cal N} := \{ T(\rho):\rho\text{ state on }\C^{d_1} \}$ is convex,
  as is the set
  ${\cal S} := \{ \tau\text{ state on }\C^{d_2}:S(\tau)\geq S(\widetilde\omega) \}$,
  and that ${\cal N}\cap{\cal S}=\{\widetilde\omega\}$. Since we assume that
  $\widetilde\omega$ is not maximally mixed, ${\cal S}$ is full--dimensional in the set of states,
  so the boundary $\partial S = \{ \tau:S(\tau)=S(\widetilde\omega) \}$ is a one--codimensional
  submanifold; from positivity of $\widetilde\omega$ (ensuring the existence of the
  derivative of $S$) it has a (unique) tangent plane $H$ at this point:
  $$ H = \Bigl\{ \xi\text{ state on }\C^{d_2}:
                 \tr\bigl[ (\xi-\widetilde\omega)\nabla S(\widetilde\omega)\bigr] = 0 \Bigr\}.$$
  Thus, $H$ is the unique hyperplane separating ${\cal S}$ from ${\cal N}$:
  \begin{align*}
    {\cal S} &\subset H^+ = \Bigl\{ \xi\text{ state on }\C^{d_2}:
                              \tr\bigl[ (\xi-\widetilde\omega)\nabla S(\widetilde\omega)\bigr]
                                                                                \geq 0 \Bigr\}, \\
    {\cal N} &\subset H^- = \Bigl\{ \xi\text{ state on }\C^{d_2}:
                              \tr\bigl[ (\xi-\widetilde\omega)\nabla S(\widetilde\omega)\bigr]
                                                                                \leq 0 \Bigr\}.
  \end{align*}
  Now consider, for phase angles $\underline\alpha=(\alpha_1,\ldots,\alpha_{d_2})$,
  the unitary
  $U_{\underline\alpha} = \sum_j e^{i\alpha_j}\ketbra{e_j}$, which clearly stabilises
  ${\cal S}$ and leaves $\widetilde\omega$ invariant. Hence, also $H$ and the
  two halfspaces $H^+$ and $H^-$ are stabilised:
  $$U_{\underline\alpha} H   U_{\underline\alpha}^* = H,  \qquad
    U_{\underline\alpha} H^+ U_{\underline\alpha}^* = H^+,\qquad
    U_{\underline\alpha} H^- U_{\underline\alpha}^* = H^-.$$
  In particular, $U_{\underline\alpha} {\cal N} U_{\underline\alpha}^* \subset H^-$, implying
  the same for the convex hull of all these sets:
  $$\conv\left\{
          \bigcup_{\underline\alpha} U_{\underline\alpha} {\cal N} U_{\underline\alpha}^*
         \right\}                                                               \subset H^-.$$
  Since this convex hull includes (for $\tau\in{\cal N}$) the states
  $$\sum_j \ketbra{e_j}\, \tau \,\ketbra{e_j}
           = \frac{1}{(2\pi)^{d_2}} \int {\rm d}\underline\alpha\, 
                                           U_{\underline\alpha} \tau U_{\underline\alpha}^*,$$
  we conclude that for all $\rho$, $\widetilde{T}(\rho)\in H^-$, forcing
  $S\bigl(\widetilde{T}(\rho)\bigr) \leq S(\widetilde\omega)$.
  \par
  We are left with the case of a degenerate $\widetilde\omega$: there we consider
  perturbations $T_\epsilon = (1-\epsilon)T+\epsilon\frac{1}{d_2}\1$ of the channel,
  whose output entropy is maximised by the same input states as $T$, and the
  optimal output state is
  $\widetilde\omega_\epsilon = (1-\epsilon)\widetilde\omega+\epsilon\frac{1}{d_2}\1$.
  These are diagonal in any diagonalising basis for $\widetilde\omega$, so
  $\widetilde{T}_\epsilon = (1-\epsilon)\widetilde{T}+\epsilon\frac{1}{d_2}\1$.
  \par
  Now our previous argument applies, and we get for all $\rho$,
  $$S\bigl( \widetilde{T}_\epsilon(\rho) \bigr) \leq S(\widetilde\omega_\epsilon)
                             \leq (1-\epsilon) S(\widetilde\omega) + \epsilon\log d_2
                                  + H(\epsilon,1-\epsilon).$$
  On the other hand, by concavity,
  $$S\bigl( \widetilde{T}_\epsilon(\rho) \bigr)
         \geq (1-\epsilon) S\bigl( \widetilde{T}(\rho) \bigr)
                                           + \epsilon\log d_2.$$
  Together, these yield for all $\rho$,
  $$S\bigl( \widetilde{T}(\rho) \bigr)
                  \leq S(\widetilde\omega) + \frac{1}{1-\epsilon}H(\epsilon,1-\epsilon),$$
  and letting $\epsilon\rightarrow 0$ concludes the proof.
\end{beweis}
\medskip
We are now in a position to prove
\begin{thm}
  \label{thm:quantum-feedback}
  For a quantum channel $T$ and $\lambda_1,\lambda_2>0$, $\lambda_1+\lambda_2<1$,
  $$\lim_{n\rightarrow\infty} \frac{1}{n}\log\log N_{\ket{F}}(n,\lambda_1,\lambda_2)
                            = C_{\rm ID}^{\ket{F}}(T) = 2 \max_\rho S\bigl(T(\rho)\bigr),$$
  unless the transmission capacity of $T$ is $0$, in which case
  $C_{\rm ID}^{\ket{F}}(T)=0$.
  \par
  In other words, the capacity of a nontrivial quantum channel
  with coherent feedback is twice
  its maximum output entropy and the strong converse holds.
\end{thm}
\begin{beweis}
  The trivial channel is easiest, and the argument is just as in
  theorem~\ref{thm:qc-feedback}. Note just one thing: a nontrivial
  channel with maximal quantum feedback
  will always allow entanglement generation (either because of the
  feedback or because it is noiseless), so --- by
  teleportation --- it will always allow quantum state transmission.
  \par
  For achievability, the sender uses $m=n-O(\log n)$ instances of the channel
  to send one half of a purification $\Psi_\rho$ of the output entropy
  maximising state $\rho$ each. This creates $m$ copies of a pure state
  which has reduced state $T(\rho)$ at the receiver.
  After performing entanglement concentration,\cite{concentration}
  which yields $n S(T(\rho)) - o(n)$ EPR pairs,
  the remaining $O(\log n)$ instances of the channel are used (with an appropriate
  error correcting code and taking some of the entanglement for
  teleportation) to implement the construction of
  proposition~\ref{prop:quantum-hashing}, based on the maximal
  entanglement.
  \par
  The converse is proved a bit differently than in theorem~\ref{thm:qc-feedback},
  where we counted the discretised decoders: now we have operators,
  and discretisation in Hilbert space is governed by slightly different rules.
  Instead, we do the following: given an identification code with feedback,
  form the uniform probabilistic mixture
  $\Phi$ of the feedback strategies $\Phi_i$ of messages $i$ --- formally,
  $\Phi=\frac{1}{N}\sum_i \Phi_i$. Its output state $\omega$ clearly
  is the uniform mixture of the output states $\omega_i$ corresponding
  to message $i$: $\omega = \frac{1}{N}\sum_i \omega_i$.
  With $\epsilon=1-\lambda_1-\lambda_2$, lemma~\ref{lemma:product}
  gives us a projector $\Pi$ of rank
  $K\leq \exp\bigl( n\max_\rho S(T(\rho)) + 48(\dim{\cal H}_2)^2\epsilon\sqrt{n} \bigr)$
  such that $\tr(\omega\Pi) \geq 1-\frac{1}{2}(\epsilon/24)^2$.
  Thus, for half of the messages (which we may assume to
  be $i=1,\ldots,\lfloor N/2 \rfloor$),
  $\tr(\omega_i\Pi)\geq 1-(\epsilon/24)^2$.
  \par
  Observe that the $\omega_i$ together with the decoding operators $D_i$
  form an identification code on ${\cal B}({\cal H}_2^{\otimes n})$,
  with error probabilities of first and second kind $\lambda_1$ and
  $\lambda_2$, respectively.
  Now restrict all $\omega_i$ and $D_i$ ($i\leq N/2$) to the supporting
  subspace of $\Pi$ (which we identify with $\C^K$):
  $$\widetilde{\omega}_i := \frac{1}{\tr(\omega_i\Pi)}\Pi\omega_i\Pi,\qquad
                                             \widetilde{D}_i := \Pi D_i \Pi.$$
  This is now an identification code on ${\cal B}(\C^K)$, with error
  probabilities of first and second kind bounded by
  $\lambda_1+\frac{1}{3}\epsilon$ and $\lambda_2+\frac{1}{3}\epsilon$,
  respectively, as a consequence of the gentle measurement
  lemma:\cite{Winter99} namely,
  $\frac{1}{2}\| \omega_i - \widetilde{\omega}_i \|_1 \leq \frac{1}{3}\epsilon$.
  So finally, we can invoke Proposition 11 of our earlier paper,\cite{winter:ID}
  which bounds the size of identification codes (this, by the way, is now the
  discretisation part of the argument):
  $$\frac{N}{2}
     \leq \left(\frac{5}{1-\lambda_1-\epsilon/3-\lambda_2-\epsilon/3}\right)^{2K^2}
     =    \left( \frac{15}{\epsilon} \right)^{2^{n \max_\rho 2S(T(\rho)) + O(\sqrt{n})}},$$
  and we have the converse.
\end{beweis}
\medskip
\begin{rem}
  \label{rem:cq-channel}
  {\rm
  For cq--channels $T:{\cal X}\longrightarrow{\cal B}({\cal H})$ (a map
  assigning a state $T(x)=\rho_x$ to every element $x$ from the finite set ${\cal X}$),
  we can even study yet another kind of feedback (let us call it
  \emph{cq--feedback}): fix purifications $\Psi_x$
  of the $\rho_x$, on ${\cal H}\otimes{\cal H}$; then input of $x\in{\cal X}$
  to the channel leads to \emph{distribution} of $\Psi_x$ between sender
  and receiver. In this way, the receiver still has the channel output state
  $\rho_x$, but is now entangled with the sender.
  \par
  By the methods employed above we can easily see
  that in this model, the identification capacity is
  $$C_{\rm ID}^{FF}(T) \geq
    \max_P \left\{ S\left( \sum_x P(x)\rho_x \right) + \sum_x P(x) S(\rho_x) \right\}.$$
  Achievability is seen as follows: for a given $P$ use a transmission
  code of rate $I(P;T) = S\bigl( \sum_x P(x)\rho_x \bigr) - \sum_x P(x) S(\rho_x)$
  and with letter frequencies $P$ in the codewords.\cite{Holevo98,SW97}
  This is used to create shared randomness of the same rate, and the
  cq--feedback to obtain pure entangled states which are
  concentrated into EPR pairs\cite{concentration}
  at rate $\sum_x P(x)E(\Psi_x) = \sum_x P(x)S(\rho_x)$;
  then we use eq.~(\ref{eq:E-and-R}).
  \par
  The (strong) converse seems to be provable by combining the approximation
  of output statistics result of Ahlswede and Winter\cite{AW}
  with a dimension counting argument as in our previous paper's\,\cite{winter:ID}
  Proposition 11, but we won't follow on this question here.
  }
\end{rem}
\medskip
\begin{rem}
  \label{rem:finalshot}
  {\rm Remarkably, the coherent feedback identification capacity
  $C_{\rm ID}^{\ket{F}}(T)$
  of a channel is at present the only one we actually ``know'' in the sense
  that we have a universally valid formula which can be evaluated
  (it is single--letter); this is in marked contrast to what
  we can say about the
  plain (non--simultaneous) identification capacity, whose determination
  remains the greatest challenge of the theory.
  }
\end{rem}

\section*{Acknowledgements}
Thanks to Noah Linden (advocatus diaboli) and
to Tobias Osborne (doctor canonicus)
for help with the proof of lemma~\ref{lemma:entropy-max}.
\par
The author was supported by the EU under European Commission
project RESQ (contract IST-2001-37559).


\begin{thebibliography}{99}\footnotesize

  \bibitem{ahlswede:GIT} R. Ahlswede, ``General Theory of Information Transfer'', 
    SFB 343 Preprint 97--118, Fakult\"at f\"ur Mathematik, Universit\"at Bielefeld (1997).

  \bibitem{ahlswede:balakirsky} R. Ahlswede, V. B. Balakirsky, ``Identification under
    Random Processes'', \emph{Probl. Inf. Transm.}, {\bf 32}(1), 123--138 (1996).

  \bibitem{AC} R. Ahlswede, I. Csisz\'{a}r, ``Common Randomness in Information Theory
    and Cryptography. I. Secret Sharing'', \emph{IEEE Trans. Inf. Theory}, {\bf 39}(4),
    1121--1132 (1993). ``Common Randomness in Information Theory
    and Cryptography. II. CR Capacity'', \emph{IEEE Trans. Inf. Theory}, {\bf 44}(1),
    225--240 (1998).

  \bibitem{AD1} R. Ahlswede, G. Dueck, ``Identification via Channels'', \emph{IEEE
    Trans. Inf. Theory}, {\bf 35}(1), 15--29 (1989).

  \bibitem{AD2} R. Ahlswede, G. Dueck, ``Identification in the Presence of Feedback
    --- a Discovery of New Capacity Formulas'', \emph{IEEE Trans. Inf. Theory},
   {\bf 35}(1), 30--36 (1989).

  \bibitem{AW} R. Ahlswede, A. Winter, ``Strong Converse for Identification via
    Quantum Channels'', \emph{IEEE Trans. Inf. Theory}, {\bf 48}(3), 569--579
    (2002). Addendum, \emph{ibid.}, {\bf 49}(1), 346 (2003).

%

  \bibitem{concentration} C. H. Bennett, H. J. Bernstein, S. Popescu, B. Schumacher,
    ``Concentrating partial entanglement by local operations'',
    \emph{Phys. Rev. A}, {\bf 53}(4), 2046--2052 (1996).

  \bibitem{rand} C. H. Bennett, P. Hayden, D. W. Leung, P. W. Shor, A. Winter,
    ``Remote preparation of quantum states'',
    \emph{e--print} {\tt quant-ph/0307100} (2003).
    P. Hayden, D. W. Leung, P. W. Shor, A. Winter, ``Randomizing quantum
    states: constructions and applications'', \emph{e--print}
    {\tt quant-ph/0307104} (2003).

%

  \bibitem{bowen} G. Bowen, ``Quantum feedback channels'',
    \emph{e--print} {\tt quant-ph/0209076} (2002).

  \bibitem{bowen:nagarajan} G. Bowen, R. Nagarajan, ``On Feedback and the Classical
    Capacity of a Noisy Quantum Channel'',
    \emph{e--print} {\tt quant-ph/0305176} (2003).

%

  \bibitem{hashing} I. Devetak, A. Winter, ``Distillation of secret key and
    entanglement from quantum states'', \emph{e--print} {\tt quant-ph/0306078}
    (2003);
    ``Relating quantum privacy and quantum coherence: an operational approach'',
    \emph{e--print} {\tt quant-ph/0307053} (2003).
    I. Devetak, A. W. Harrow, A. Winter, ``A family of quantum protocols'',
    \emph{e--print} {\tt quant-ph/0308044} (2003).


  \bibitem{fujiwara:nagaoka} A. Fujiwara, H. Nagaoka, ``Operational capacity and
    pseudoclassicality of a quantum channel'',
    \emph{IEEE Trans. Inf. Theory}, {\bf 44}(3), 1071--1086 (1998).


  \bibitem{han:verdu} T. S. Han, S. Verd\'{u}, ``Approximation theory of output
    statistics'', \emph{IEEE Trans. Inf. Theory}, {\bf 39}(3), 752--772 (1993).

  \bibitem{ccc} A. W. Harrow, ``Coherent Communication of Classical Messages'',
    \emph{Phys. Rev. Lett.}, {\bf 92}(9), 097902 (2004).

  \bibitem{quantum-SD} A. Harrow, P. Hayden, D. W. Leung, ``Superdense coding
    of quantum states'', \emph{e--print} {\tt quant-ph/0307221} (2003).


  \bibitem{Holevo77} A. S. Holevo, ``Problems in the mathematical theory of quantum
    communication channels'', \emph{Rep. Math. Phys.}, {\bf 12}(2), 273--278 (1977).


  \bibitem{Holevo98} A. S. Holevo, ``The capacity of the quantum channel with
    general signal states'', \emph{IEEE Trans. Inf. Theory}, {\bf 44}(1),
    269--273 (1998).


  \bibitem{jamiolkowski} A. Jamio\l{}kowski, ``Linear transformations which preserve
    trace and positive semidefiniteness of operators'', \emph{Rep. Math. Phys.},
    {\bf 3}, 275--278 (1972).

  \bibitem{Kleinewaechter} C. Kleinew\"achter, \emph{On Identification},
    Ph.D. thesis, University of Bielefeld. Unpublished.
    SFB 343 Preprint 99--064, Fakult\"at f\"ur Mathematik, Universit\"at Bielefeld
    (1999). Online at {\tt www.mathematik.uni-bielefeld.de/sfb343/preprints/}


  \bibitem{kuperberg} G. Kuperberg, ``The capacity of hybrid quantum memory'',
    \emph{IEEE Trans. Inf. Theory}, {\bf 49}(6), 1465--1473 (2003).


  \bibitem{Loeber} P. L\"ober, \emph{Quantum Channels and Simultaneous ID Coding},
    Ph.D. thesis, Universit\"at Bielefeld, Bielefeld, Germany (1999). Unpublished.
    Available as \emph{e--print} {\tt quant-ph/9907019}.

  \bibitem{random:states} E. Lubkin, ``Entropy of an $n$--system from its correlation
    with a $k$--reservoir'', \emph{J. Math. Phys.}, {\bf 19}, 1028--1031 (1978).
    D. Page, ``Average entropy of a subsystem'', \emph{Phys. Rev. Lett.}, {\bf 71}(9),
    1291--1294 (1993).


  \bibitem{Ogawa:Nagaoka} T. Ogawa, H. Nagaoka, ``Strong Converse to the Quantum
    Channel Coding Theorem'', \emph{IEEE Trans. Inf. Theory}, {\bf 45}(7),
    2486--2489 (1999).

%

  \bibitem{SW97} B. Schumacher, M. D. Westmoreland, ``Sending classical information
    via noisy quantum channels'', \emph{Phys. Rev. A}, {\bf 56}(1), 131--138 (1997).

  \bibitem{Shannon} C. E. Shannon, ``A mathematical theory of communication'',
    \emph{Bell System Tech. J.}, {\bf 27}, 379--423 and 623--656 (1948).

  \bibitem{shor:add} P. W. Shor, ``Equivalence of Additivity Questions in
    Quantum Information Theory'', \emph{e--print} {\tt quant-ph/0305035}.
    To appear in \emph{Comm. Math. Phys.} (2004).


  \bibitem{steinberg:merhav} Y. Steinberg, N. Merhav, ``Identification in the Presence
    of Side Information with Applications to Watermarking'',
    \emph{IEEE Trans. Inf. Theory}, {\bf 47}(4), 1410--1422 (2001).

  \bibitem{Winter99} A. Winter, ``Coding Theorem and Strong Converse for Quantum
    Channels'', \emph{IEEE Trans. Inf. Theory}, {\bf 45}(7), 2481--2485 (1999).

%

  \bibitem{winter:ID} A. Winter ``Quantum and Classical Message Identification
    via Quantum Channels'', \emph{e--print} {quant-ph/0401060} (2004).

%

\end{thebibliography}
\end{document}